\pdfoutput=1
\newcommand{\PreprintNote}{Presented at the 161st AES Convention,
  Nashville, TN, USA, October 30--November 1, 2026.
  Published version: \url{https://aes.org/publications/elibrary/elibrary-page/?id=23480}}
\documentclass{preprint} %

\graphicspath{{./}{figures/}}

\usepackage[utf8]{inputenc}

\usepackage{amsmath}

\usepackage{microtype}
\makeatletter
\providecommand\MT@suspend@tagging{}
\providecommand\MT@resume@tagging{}
\makeatother

\usepackage[numbers,square]{natbib}

\usepackage{booktabs}
\usepackage{color}
\usepackage{url}
\usepackage{tikz}
\usetikzlibrary{positioning, arrows.meta, calc}

\usepackage{caption}
\usetikzlibrary{positioning, arrows.meta, calc}

\tikzset{
    block/.style={
        draw, rectangle, minimum height=3.2em, minimum width=7.5em,
        text centered, text width=6.5em, font=\small, fill=white, line width=0.8pt,
        inner sep=4pt, align=center
    },
    wide block/.style={block, minimum width=8.5em, text width=7.5em},
    arrow/.style={-{Stealth[length=6pt]}, line width=0.8pt},
    dashed arrow/.style={arrow, dashed},
    fork/.style={fill=black, circle, inner sep=1.5pt},
}

\title{Lightweight Neural Prediction for MDCT-Based Perceptual Audio Coding}

\author[1]{Niccolo Abate}
\author[1]{Gabriel Soule}
\author[1]{Marina Bosi}

\affil[1]{Center for Computer Research in Music and Acoustics (CCRMA), Stanford University, Stanford, CA, USA}

\correspondence{Niccolo Abate}{nlabate@stanford.edu}

\lastnames{Abate, Soule, and Bosi}

\shorttitle{Neural MDCT Prediction}

\begin{document}

\twocolumn[
\maketitle %

\begin{onecolabstract}
  Perceptual audio codecs often use \textit{predictive coding} to further reduce signal redundancy. Rather than learning the entire codec pipeline end-to-end, we insert a small neural network as a drop-in, deterministic predictor in an otherwise conventional codec: the network estimates the next Modified Discrete Cosine Transform (MDCT) coefficient block from a short context of reconstructed blocks, and the encoder transmits the quantized residual. We compare Multilayer Perceptron (MLP) and Convolutional Neural Network (CNN) predictors against a non-predictive baseline and a per-line Normalized Least Mean Squares (NLMS) linear predictor (similar to the MPEG-2 AAC Main Profile predictor). The CNN consistently outperforms the MLP, the linear predictor, and the baseline in signal-to-noise ratio (SNR), noise-to-mask ratio (NMR), and MUSHRA evaluation scores. Using a 512-sample analysis window ($w=512$), our neural predictors recover, and the CNN exceeds, the coded quality obtained by the non-predictive baseline after quadrupling the analysis-window length to $w=2048$, at substantially lower algorithmic delay. The CNN uses only $0.13$M parameters (roughly $20\times$ fewer than the MLP) and has a decoder real-time factor (RTF) of $\approx 0.10$ on a consumer CPU.
\end{onecolabstract}
]

\section{Introduction}
A perceptual audio codec~\citep{bosi_introduction_2003} is designed to encode audio at the lowest possible bitrate while maintaining high perceived audio quality for the listener.

Audio signals are rarely uniformly random; they carry structure across both time and frequency. A \textit{predictor} exploits this structure by estimating the next value in a sequence from a window of previous samples, and if effective can be integrated into a codec pipeline to improve coding efficiency.

Predictive coding is a well-appreciated technique within popular codecs. MPEG-4 AAC \citep{isoiec_aac_2019} uses both a backward-adaptive linear predictor on MDCT coefficients and a forward-adaptive long-term predictor, while Opus \citep{valin_opus_2012} applies LPC and pitch prediction in its SILK speech path and a first-order envelope predictor across frames in its CELT music path. For many telephony applications, low codec delay is a necessary feature to ensure natural and interactive communication. For example, the AAC-LD~\citep{allamanche_mpeg4_1999} and AAC-ELD~\cite{schnell_mpeg4_2008} profiles provide low-delay options by shortening the MDCT window size, reducing algorithmic delay at the cost of coarser frequency resolution.

Rather than learning the entire codec pipeline end-to-end, we propose a new approach in which a small neural network acts as a drop-in, deterministic predictor in an otherwise conventional codec. We implement both an MLP and a CNN as small, portable predictors and compare them to a non-predictive baseline implementation as well as a linear predictor similar to the one used in MPEG-2 AAC Main. We show that the MLP and CNN outperform both the linear predictor and the non-predictive baseline, with respect to prediction error, SNR, NMR, and perceptual quality (via a MUSHRA listening test). We show that, using a 512-sample analysis window ($w=512$), our neural predictors recover, and the CNN exceeds, the coded quality obtained by the non-predictive baseline after quadrupling the MDCT size to $w=2048$, at substantially lower algorithmic delay.

\section{Prior Art}
Recently, deep neural networks have been employed with increasing frequency in audio coding. The most prominent line of work is the fully end-to-end neural codec: systems such as SoundStream~\citep{zeghidour_soundstream_2022}, EnCodec~\citep{defossez_high_2023}, and the Descript Audio Codec~\citep{kumar_high-fidelity_2023} learn the encoder, quantizer, and decoder jointly, pairing a convolutional autoencoder with residual vector quantization and adversarial training to reconstruct audio at very low bitrates. These models reach impressive quality, but they replace many parts of the conventional codec pipeline with a learned black box, which complicates integration with existing transform-domain codecs and standards. A recent survey further observes that such purely data-driven codecs often require large architectures merely to match the efficiency of established model-based coders, and identifies \emph{hybrid} designs, which combine learned components with conventional coding, as a promising middle ground \citep{kim_neural_2024}.

A complementary body of work keeps much of the conventional coder intact and introduces neural techniques in a modular fashion. Several systems operate directly on transform-domain coefficients: MDCTNet is a recurrent generative model that predicts MDCT coefficients in the decoder, conditioned on a perceptually coded representation \citep{davidson_high_2022}, while other recent codecs operate on the MDCT spectrum but learn the surrounding quantization and coding components \citep{lim_perceptual_2024,jiang_mdctcodec_2024}. Others insert a neural module for a single stage of an otherwise standard codec---for example replacing spectral band replication with a learned bandwidth-extension network on top of a fixed AAC core \citep{choi_neural_2025}---or target lightweight, streaming operation explicitly \citep{ahn_hilcodec_2024}. Closest to our setting, neural \emph{prediction} has been used to remove temporal redundancy before coding: TF-Codec predicts a learned latent representation from past quantized frames \citep{jiang_latent-domain_2023}, and \citet{yang_neural_2023} pair a recurrent feature predictor with discriminative residual coding. These predictive schemes, however, operate on learned speech features and remain coupled to neural quantizers or vocoders, rather than acting as a drop-in module inside a standard transform coder.

In this work, we study neural prediction as a generic, drop-in component for an existing codec pipeline: a deterministic predictor is inserted into a conventional MDCT residual-coding path while the transform, psychoacoustic analysis, quantizer, and synthesis stages are left untouched and entirely independent of the predictor. We are particularly interested in small models that may be evaluated quickly and in low-latency contexts.

\section{Methods}

Our experimental codec is built on top of a conventional MDCT-based perceptual audio coder \cite{bosi_introduction_2003}. The codec applies an MDCT~\citep{1164954} with a Kaiser--Bessel-derived (KBD) window~\citep{Fielder1996ac-2} to overlapping $w=2K$-sample analysis windows, producing one $K$-line MDCT coefficient block per audio channel every $K$-sample hop.

The MDCT lines are grouped into scale-factor bands corresponding to critical-band partitions \cite{zwicker_psychoacoustics_2007}. For each coefficient block, the encoder computes signal-to-mask ratios (SMRs) for each band based on psychoacoustic models~\cite{bosi_introduction_2003}. Based on the SMRs and the target bitrate constraint, the bit budget is allocated to optimize perceived audio quality.

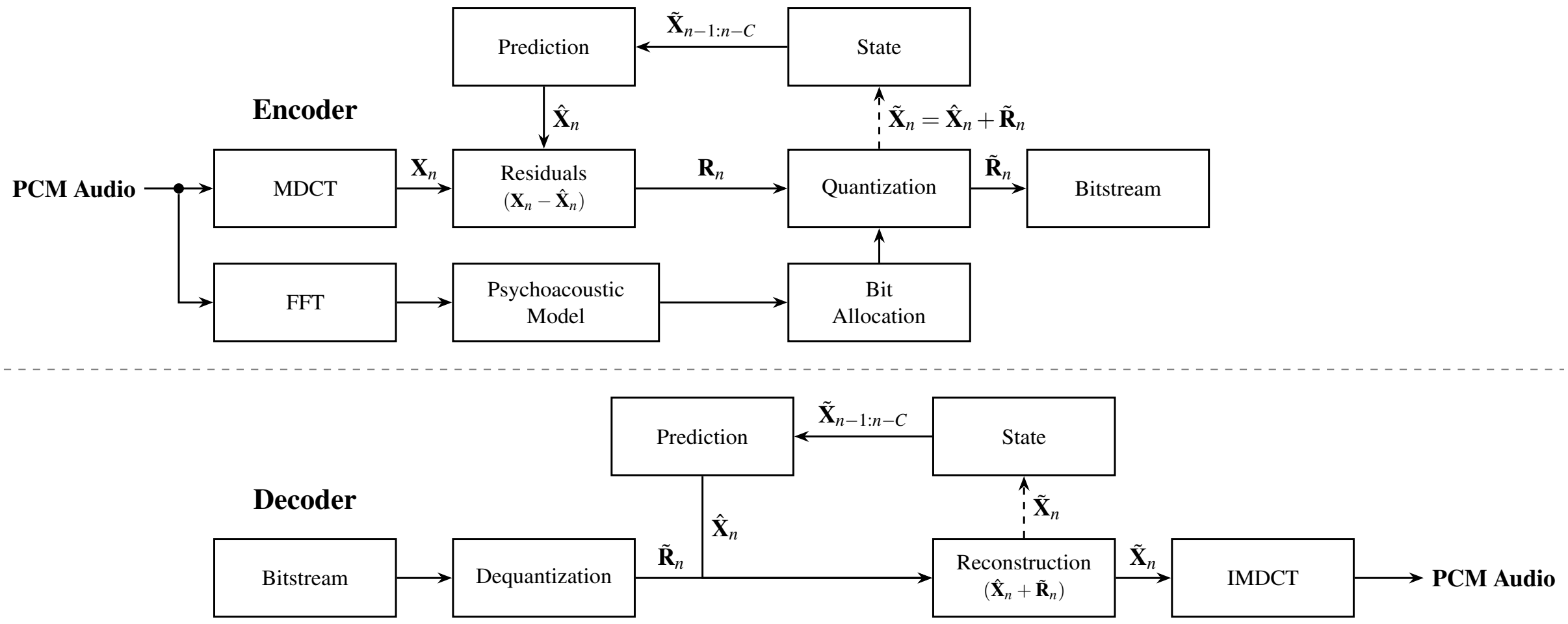
\begin{figure*}[t]
\centering
\resizebox{\textwidth}{!}{%
\begin{tikzpicture}[node distance=1.0cm and 0.8cm]

    \node[font=\large\bfseries] (enc_title) {Encoder};

    \node[block, below=0.3cm of enc_title] (mdct) {MDCT};
    \node[block, right=of mdct] (res) {Residuals\\{\footnotesize$(\mathbf{X}_n - \hat{\mathbf{X}}_n)$}};
    \node[block, right=2.2cm of res] (quant) {Quantization};
    \node[block, right=of quant] (enc_bits) {Bitstream};

    \node[block, above=0.9cm of res] (enc_pred) {Prediction};

    \draw[arrow] (mdct) -- node[above] {$\mathbf{X}_n$} (res);
    \draw[arrow] (enc_pred) -- node[right] {$\hat{\mathbf{X}}_n$} (res);
    \draw[arrow] (res) -- node[above] {$\mathbf{R}_n$} (quant);
    \draw[arrow] (quant) -- node[above] {$\tilde{\mathbf{R}}_n$} (enc_bits);

    \node[left=1.0cm of mdct] (input) {\textbf{PCM Audio}};
    \draw[arrow] (input) -- (mdct);

    \node[fork] at ($(input.east)!0.5!(mdct.west)$) (fk) {};

    \node[block, below=0.5cm of mdct] (fft) {FFT};
    \node[wide block, right=of fft] (psy) {Psychoacoustic\\Model};
    \node[block, at={(quant |- fft)}] (bitalloc) {Bit\\Allocation};

    \draw[arrow] (fk) |- (fft);
    \draw[arrow] (fft) -- (psy);
    \draw[arrow] (psy) -- (bitalloc);
    \draw[arrow] (bitalloc.north) -- (quant.south);

    \node[block, above=0.9cm of quant] (enc_state) {State};

    \draw[dashed arrow] (quant.north) -- node[right] {$\tilde{\mathbf{X}}_n = \hat{\mathbf{X}}_n + \tilde{\mathbf{R}}_n$} (enc_state.south);
    \draw[arrow] (enc_state.west) -- node[above] {$\tilde{\mathbf{X}}_{n-1:n-C}$} (enc_pred.east);

    \node[font=\large\bfseries, below=2.0cm of fft] (dec_title) {Decoder};

    \node[block, below=0.3cm of dec_title] (dec_bits) {Bitstream};
    \node[block, right=of dec_bits] (dequant) {Dequantization};
    \node[block, right=4.3cm of dequant] (recon) {Reconstruction\\{\footnotesize$(\hat{\mathbf{X}}_n + \tilde{\mathbf{R}}_n)$}};
    \node[block, right=of recon] (imdct) {IMDCT};

    \node[block, above=0.9cm of recon] (dec_state) {State};
    \node[block, left=2.0cm of dec_state] (dec_pred) {Prediction};

    \draw[arrow] (dec_bits) -- (dequant);
    \draw[arrow] (dequant) -- node[above, pos=0.12] {$\tilde{\mathbf{R}}_n$} (recon);
    \draw[arrow] (dec_pred) |- node[near start, right] {$\hat{\mathbf{X}}_n$} (recon);
    \draw[arrow] (recon) -- node[above] {$\tilde{\mathbf{X}}_n$} (imdct);

    \node[right=1.0cm of imdct] (output) {\textbf{PCM Audio}};
    \draw[arrow] (imdct) -- (output);

    \draw[dashed arrow] (recon.north) -- node[right] {$\tilde{\mathbf{X}}_n$} (dec_state.south);
    \draw[arrow] (dec_state.west) -- node[above] {$\tilde{\mathbf{X}}_{n-1:n-C}$} (dec_pred.east);

    \path (fft.south) -- (dec_state.north) coordinate[midway] (sep_mid);
    \path let \p1 = (input.west), \p2 = (dec_bits.west) in
        coordinate (sep_left) at ({min(\x1,\x2)}, 0);
    \path let \p1 = (enc_bits.east), \p2 = (output.east) in
        coordinate (sep_right) at ({max(\x1,\x2)}, 0);
    \draw[dashed, gray, line width=0.5pt]
        (sep_left |- sep_mid) -- (sep_right |- sep_mid);

\end{tikzpicture}}
\caption{Block diagram of the predictive codec. In flagged scale-factor bands, the encoder transmits quantized residual coefficients; unflagged bands carry quantized original coefficients. An identical decoder-side predictor instance reproduces $\hat{\mathbf{X}}_n$ from previously reconstructed coefficient blocks $\tilde{\mathbf{X}}_{n-1:n-C}$.}
\label{fig:pipeline}
\end{figure*}

The predictive codec inserts a deterministic predictor between MDCT analysis and quantization (Fig.~\ref{fig:pipeline}). Let $\mathbf{X}_n$ denote the MDCT coefficient block at block index $n$, and let $\tilde{\mathbf{X}}_n$ denote its decoder-equivalent reconstruction. Given a context of the previous $C$ reconstructed blocks, the predictor estimates the current block as
\begin{equation}
    \label{eq:predict}
    \hat{\mathbf{X}}_n = P_\theta(\tilde{\mathbf{X}}_{n-C}, \ldots, \tilde{\mathbf{X}}_{n-1}).
\end{equation}
The encoder forms a residual $\mathbf{R}_n = \mathbf{X}_n - \hat{\mathbf{X}}_n$ and may quantize this residual instead of the original MDCT coefficients. At the decoder, the predictor produces $\hat{\mathbf{X}}_n$ from previously reconstructed blocks and adds it back in prediction-enabled bands to obtain $\tilde{\mathbf{X}}_n$. During the initial blocks, before enough context is available, the codec falls back to the non-predictive MDCT path.

Prediction is applied conditionally at the scale-factor-band level: for each band the encoder transmits a one-bit flag selecting between the residual and baseline candidates according to their provisional reconstruction errors, reducing exposure to harmful predictions.

Because the predictor state is updated only from reconstructed MDCT coefficient blocks --- the same dequantized coefficients the decoder will see --- both sides remain synchronized without transmitting predictor state. The predictor can therefore be treated as a black-box module, provided it is deterministic and exposes the same predict/update behavior at the encoder and decoder.

\subsection{Neural Predictor Architecture}

We implement both MLP and CNN predictors. Both predictors consume a context of $C$ reconstructed MDCT coefficient blocks, each with $K$ lines, and emit a single $K$-line MDCT estimate, trained to minimize the mean-squared error between the predicted and target blocks.

The MLP predictor flattens the context into a length-$KC$ vector and passes it through two fully-connected hidden layers of width 1024, each followed by batch normalization, a ReLU nonlinearity, and dropout. A linear output layer produces the $K$-line prediction.

The CNN predictor treats the $C$ context blocks as input channels of a 1D convolution over the frequency axis, yielding a $C \times K$ tensor. The stem applies $64$ filters of shape $C \times 5$ --- each filter spans all context blocks at $5$ adjacent frequency bins --- and slides them across $K$, followed by batch normalization, ReLU, and dropout. Three residual blocks, each two such convolutions deep, extend the receptive field across neighboring MDCT lines, and a final pointwise convolution collapses the channel axis to produce the $K$-line prediction.

The two architectures trade flexibility for built-in assumptions in opposite ways. The MLP is fully connected, so every output line can depend on every input line in every context block. This gives it the most general possible mapping, but no built-in priors, such as locality or translation equivariance. The CNN is constrained by construction: each output line is computed from only a limited window of neighboring lines, and the same filter is applied at every frequency. These two constraints match the structure of MDCT spectra, where related energy tends to occupy adjacent lines, and similar local patterns recur at different frequencies, suggesting a good fit for a CNN.

\subsection{Linear Predictor Baseline}

As a non-neural point of comparison we implement a per-line second-order backward-adaptive linear predictor, in the spirit of the backward-adaptive linear prediction introduced
in MPEG-2 AAC~\citep{bosi_introduction_2003,isoiec_aac_2019,Bosi1997MPEG2AAC}.%

Each MDCT line is predicted independently from its own two most recent reconstructed values $r_1 = \tilde{X}_{n-1}[k]$ and $r_2 = \tilde{X}_{n-2}[k]$, exploiting the temporal correlation between corresponding MDCT-line values in successive blocks:
\begin{equation}
    \label{eq:linpred}
    \begin{aligned}
        \hat{X}_n[k] &= a_1(n)\, r_1 + a_2(n)\, r_2, \\
        e(n) &= \tilde{X}_n[k] - \hat{X}_n[k].
    \end{aligned}
\end{equation}
The per-line coefficients are updated via normalized least-mean-squares \citep{haykin_adaptive_1996}:
\begin{equation}
    \label{eq:nlms}
    a_i(n+1) = \rho\, a_i(n) \;+\; \mu\, \frac{e(n)\, r_i}{r_1^2 + r_2^2 + \epsilon}, \quad i \in \{1, 2\},
\end{equation}
with step size $\mu = 0.1$, leakage $\rho = 0.995$, and a small $\epsilon$ regularizing the denominator, and clamped to $a_1 \in [-1.9, 1.9]$, $a_2 \in [-0.95, 0.95]$

\subsection{Conditional Prediction Gating and Bit Allocation}

Predictors will sometimes predict incorrectly. Our codec therefore uses a quantization-aware gate to reduce the risk that residual coding increases coefficient-domain reconstruction error. Fix a block index $n$ and scale-factor band $b$, and let $\mathcal{Q}(\mathbf{Y})$ denote the codec's quantize--dequantize round trip applied to the entire coefficient block $\mathbf{Y}$ under the current bit allocation, including the overall scale selected from $\mathbf{Y}$. With $\mathbf{R}=\mathbf{X}-\hat{\mathbf{X}}$, the encoder evaluates all-original and all-residual candidates:
\begin{equation}
    \label{eq:gate}
\begin{aligned}
    \varepsilon^{\mathrm{base}}_b &= \bigl\| \mathbf{X}^{(b)} - [\mathcal{Q}(\mathbf{X})]^{(b)} \bigr\|^2, \\
    \varepsilon^{\mathrm{pred}}_b &= \bigl\| \mathbf{X}^{(b)} - \bigl(\hat{\mathbf{X}}^{(b)} + [\mathcal{Q}(\mathbf{R})]^{(b)}\bigr) \bigr\|^2,
\end{aligned}
\end{equation}
and sets band $b$'s prediction flag iff $\varepsilon^{\mathrm{pred}}_b < \varepsilon^{\mathrm{base}}_b$.

These candidate round trips account for quantizer behavior omitted by a comparison of unquantized band energies. They are nevertheless provisional: after mask selection, the encoder assembles a mixed coefficient block containing original coefficients in unflagged bands and residuals in flagged bands, then quantizes it again, including recomputation of the block-wide and per-band scale factors. The final decoder errors can therefore differ from the candidate errors, though this difference is minor in practice.

The flags are interleaved with the bit-allocation payload, one raw bit per scale-factor band, and the decoder adds the prediction only in flagged bands. Here, the baseline candidate uses the masked format's bit allocation; the signaling overhead precludes a strict rate-matched guarantee against the no-mask codec.

\subsection{Training and Evaluation Methodology}

The neural predictors are trained on Medley-solos-DB \citep{lostanlen_2019_3464194}, a cross-collection corpus of solo musical-instrument recordings at 44.1~kHz. Each recording is analyzed with the same KBD-windowed MDCT used by the codec, and the resulting coefficient blocks are normalized by a single global standard deviation estimated over the corpus, to bring them into a more tractable range for learning. A random $10\%$ of files is held out for validation during training.

Each predictor consumes a context of $C$ previously reconstructed blocks, and we train models over a range of context lengths. During training, no quantization is applied; the original MDCT blocks are used, equivalent to a perfectly reconstructed context rather than the codec's quantized one. Training minimizes the mean-squared error against the next block
using the Adam optimizer (learning rate $10^{-4}$), for up to $1000$ epochs. The network predicts the increment over the most recent reconstructed block, $\hat{\mathbf{X}}_n = \tilde{\mathbf{X}}_{n-1} + f_\theta(\tilde{\mathbf{X}}_{n-C}, \ldots, \tilde{\mathbf{X}}_{n-1})$.

We evaluate the complete codec on the $70$-item EBU SQAM test set \citep{ebu_sqam_2008}, which spans single instruments, vocals, speech, ensembles, and artificial signals; because this corpus is entirely disjoint from the Medley-solos-DB training data, all reported results measure cross-corpus generalization rather than in-distribution fit. Unless stated otherwise we report predictors with $C = 5$ context blocks at $2.0$~bits/sample. We report two analysis-window lengths: $w=512$ samples ($K=256$ MDCT lines) and $w=2048$ samples ($K=1024$ lines). In every configuration the two neural predictors (the MLP and the CNN) are compared against the non-predictive baseline and the NLMS linear predictor using the same transform, psychoacoustic model, and bit-allocation procedure; predictive modes additionally reserve one flag bit per scale-factor band.

We report three objective measures: prediction gain, reconstruction SNR, and the noise-to-mask ratio (NMR), together with a subjective MUSHRA listening test following the ITU-R BS.1534 \citep{itur_bs1534_2015} recommendation. \emph{Prediction gain} (dB) characterizes a predictor in isolation as the ratio of target-block energy to residual energy. %
While SNR does not account for masking and is thus an imperfect proxy for perceptual quality on its own, reconstruction \emph{SNR} (dB), or \emph{coded SNR}, is a meaningful comparative measure across predictors that share an identical perceptual codec. The \emph{NMR} (computed as SMR $-$ SNR, per scale-factor band) places the quantization noise relative to the masking threshold.

\section{Results}

\subsection{Objective Quality: SNR and NMR}
\label{sec:objective}

Table~\ref{tab:objective_means} shows that the predictors order monotonically across mean prediction gain, reconstruction SNR, and NMR as follows: Baseline (no predictor) $<$ NLMS linear $<$ MLP $<$ CNN, with the CNN clearly preferred in all metrics. The MLP beats the linear predictor on SNR but ties it on mean NMR, making the CNN the only learned predictor with a clear masking-aware advantage over linear ($\approx 1.1$~dB lower mean NMR).

\begin{table}[t]
\caption{Mean prediction gain (Gain), SNR, and NMR (all dB) across the numbered 70-file EBU SQAM set at $2.0$~bits/sample, $w=512$, $C=5$. $\Delta$ values are relative to the non-predictive baseline; %
higher Gain and SNR values indicate better performance, whereas lower (more negative) NMR values indicate better performance.}
\begin{center}
\begin{small}
\setlength{\tabcolsep}{4pt}
\begin{tabular}{lccccc}
\toprule
Mode     & Gain   & SNR     & $\Delta$ SNR & NMR      & $\Delta$ NMR \\
\midrule
Baseline & ---    & $18.86$ & ---          & $-15.56$ & ---          \\
Linear   & $6.38$ & $21.35$ & $+2.49$      & $-18.19$ & $-2.63$      \\
MLP      & $8.97$ & $22.44$ & $+3.58$      & $-18.23$ & $-2.67$      \\
CNN      & $10.06$ & $23.18$ & $+4.32$      & $-19.27$ & $-3.71$      \\
\bottomrule
\end{tabular}
\end{small}
\label{tab:objective_means}
\end{center}
\end{table}

Looking at each individual critical item,
this trend is even more pronounced (Tables~\ref{tab:perfile_snr} and~\ref{tab:perfile_nmr}).
Raw prediction gains shrink substantially through the codec's quantization and bit allocation, yet even the NLMS linear predictor retains a sizable baseline-to-prediction jump. As expected, the largest coded gains come on tonal, harmonically rich content, where consecutive blocks are strongly correlated; the handful of items on which prediction \emph{hurts} cluster on transient and stochastic content (percussion, broadband noise). On cello, flute, and soprano the CNN reaches $+7$~dB SNR over baseline and roughly $+3$--$4$~dB over the linear predictor, with the MLP just behind at roughly $+6$~dB and $+2$--$3$~dB respectively. On the plucked and struck items (harpsichord and glockenspiel respectively), the linear predictor matches or beats the MLP, but the CNN re-establishes its lead.

\begin{table}[t]
\caption{Mean SNR (dB) per critical item, same configuration as Table~\ref{tab:objective_means}.}
\begin{center}
\begin{small}
\begin{tabular}{lcccc}
\toprule
File                          & Base.  & Linear & MLP    & CNN             \\
\midrule
Glockenspiel                  & $16.1$ & $19.3$ & $18.5$ & $\mathbf{21.3}$ \\
Harpsichord                   & $11.1$ & $12.5$ & $12.6$ & $\mathbf{14.0}$ \\
Organ                         & $11.5$ & $13.9$ & $15.4$ & $\mathbf{16.3}$ \\
Soprano                       & $18.3$ & $22.0$ & $24.8$ & $\mathbf{25.8}$ \\
Male (EN)                     & $17.0$ & $17.9$ & $\mathbf{19.1}$ & $19.1$ \\
Cello                         & $17.8$ & $21.6$ & $23.3$ & $\mathbf{24.4}$ \\
Flute                         & $16.2$ & $19.6$ & $23.3$ & $\mathbf{23.8}$ \\
\midrule
\textbf{Mean (70 files)}  & $18.9$ & $21.4$ & $22.4$ & $\mathbf{23.2}$ \\
\bottomrule
\end{tabular}
\end{small}
\label{tab:perfile_snr}
\end{center}
\end{table}

\begin{table}[t]
\caption{Mean NMR (dB) per scale-factor band, averaged per critical item. Same configuration as Table~\ref{tab:objective_means}.}
\begin{center}
\begin{small}
\begin{tabular}{lcccc}
\toprule
File                          & Base.   & Linear  & MLP     & CNN              \\
\midrule
Glockenspiel                  & $-19.1$ & $-21.8$ & $-21.0$ & $\mathbf{-22.6}$ \\
Harpsichord                   & $-12.1$ & $-15.6$ & $-14.9$ & $\mathbf{-17.3}$ \\
Organ                         & $-9.3$  & $-13.2$ & $-13.3$ & $\mathbf{-15.0}$ \\
Soprano                       & $-12.9$ & $-15.7$ & $-16.4$ & $\mathbf{-17.8}$ \\
Male (EN)                     & $-12.1$ & $-13.6$ & $-14.0$ & $\mathbf{-14.4}$ \\
Cello                         & $-11.3$ & $-14.5$ & $-15.1$ & $\mathbf{-16.6}$ \\
Flute                         & $-10.1$ & $-13.9$ & $-14.4$ & $\mathbf{-16.0}$ \\
\midrule
\textbf{Mean (70 files)}  & $-15.6$ & $-18.2$ & $-18.2$ & $\mathbf{-19.3}$ \\
\bottomrule
\end{tabular}
\end{small}
\label{tab:perfile_nmr}
\end{center}
\end{table}

The CNN produces more consistent results, outperforming the linear predictor in raw prediction gain and outperforming both the baseline and linear predictor in SNR and NMR, whereas the MLP approaches the linear predictor in mean NMR and in the per-item results. The CNN's priors push it toward general spectral structure, while the MLP's unconstrained capacity is free to fit patterns specific to the training corpus that do not transfer,\footnote{Scaling the MLP to three hidden layers ($3.69$M parameters) lowers its validation loss on held-out Medley-solos-DB files by $17\%$ --- the best of the three models in-distribution --- yet raises cross-corpus prediction gain on the SQAM set by only $0.07$~dB ($8.97 \rightarrow 9.04$~dB), still about $1$~dB behind the CNN.} so here the $0.13$M-parameter CNN outperforms the $2.63$M-parameter MLP despite its far smaller size.

\subsection{Latency Considerations}
\label{sec:blocksize}

For our 50\%-overlapped MDCT implementation, the end-to-end algorithmic delay comprises one $K$-sample input hop of buffering plus the $K$-sample analysis/synthesis filter-bank delay, for a total of $w=2K$ samples: $11.6$~ms at $w=512$ versus $46.4$~ms at $w=2048$ ($f_s = 44.1$~kHz). Low-delay coders such as MPEG-4 AAC-LD~\citep{allamanche_mpeg4_1999} and AAC-ELD~\citep{schnell_mpeg4_2008} operate between these, using $w=1024$ paired with other low-delay techniques\footnote{AAC-ELD's LD-MDCT lowers the filter-bank delay by shortening the window's overlap toward future values; our codec could adopt it to reduce delay further.} to reach $\sim$15--30~ms algorithmic delay at 48~kHz. The shorter MDCT window provides coarser frequency resolution, however, yielding a weaker non-predictive baseline; we therefore study prediction at both analysis-window lengths, asking primarily whether prediction at $w=512$ can recover the quality of the $w=2048$ configuration.

\begin{table}[t]
\caption{Mean prediction gain (Gain), SNR, and NMR (dB) at $2.0$~bits/sample, comparing $w=512$ ($C=5$) against $w=2048$ ($C=3$; the $w=2048$ hop spans $4\times$ as many samples as the $w=512$ hop, so $C=3$ gives the same look-back duration as $C=12$ at $w=512$).}
\begin{center}
\begin{small}
\setlength{\tabcolsep}{3.5pt}
\begin{tabular}{lcccccc}
\toprule
         & \multicolumn{3}{c}{$w=512$} & \multicolumn{3}{c}{$w=2048$} \\
\cmidrule(lr){2-4} \cmidrule(lr){5-7}
Mode     & Gain   & SNR     & NMR      & Gain   & SNR     & NMR      \\
\midrule
Baseline & ---    & $18.86$ & $-15.56$ & ---    & $21.93$ & $-18.11$ \\
Linear   & $6.38$ & $21.35$ & $-18.19$ & $3.00$ & $23.84$ & $-19.70$ \\
MLP      & $8.97$ & $22.44$ & $-18.23$ & $4.84$ & $23.99$ & $-19.20$ \\
CNN      & $10.06$ & $23.18$ & $-19.27$ & $8.01$ & $24.98$ & $-20.47$ \\
\bottomrule
\end{tabular}
\end{small}
\label{tab:blocksize_means}
\end{center}
\end{table}

As expected, increasing the analysis-window length to $w=2048$ improves the non-predictive baseline (Table~\ref{tab:blocksize_means}). Even so, the $w=512$ MLP and CNN both exceed the $w=2048$ baseline in coded SNR (CNN $+1.2$~dB, despite using one quarter of the analysis-window length), with the $w=512$ CNN trailing only the predictor-equipped $w=2048$ configurations. In NMR, the simple $w=512$ linear predictor already matches the $w=2048$ baseline, and the $w=512$ CNN lands between the $w=2048$ linear and MLP predictors. %
At $w=2048$, the CNN gains $+3.05$~dB SNR and $-2.36$~dB NMR over the non-predictive baseline. The per-item perspective supports similar conclusions (Table~\ref{tab:blocksize_perfile}): on $5$ of $7$ items the $w=512$ MLP meets or exceeds the $w=2048$ baseline in SNR, and the $w=512$ CNN does so on $6$ of $7$.
\begin{table}[t]
\caption{Mean SNR (dB) per file: $w=2048$ baseline against the four $w=512$ configurations. Bolded $w=512$ cells meet or exceed the $w=2048$ baseline.}
\begin{center}
\begin{small}
\setlength{\tabcolsep}{3pt}
\begin{tabular}{lccccc}
\toprule
         & $w=2048$ & \multicolumn{4}{c}{$w=512$} \\
\cmidrule(lr){3-6}
File                          & Base    & Base    & Linear           & MLP              & CNN              \\
\midrule
Glockenspiel                  & $19.7$  & $16.1$  & $19.3$           & $18.5$           & $\mathbf{21.3}$  \\
Harpsichord                   & $11.3$  & $11.1$  & $\mathbf{12.5}$  & $\mathbf{12.6}$  & $\mathbf{14.0}$  \\
Organ                         & $13.7$  & $11.5$  & $\mathbf{13.9}$  & $\mathbf{15.4}$  & $\mathbf{16.3}$  \\
Soprano                       & $23.5$  & $18.3$  & $22.0$           & $\mathbf{24.8}$  & $\mathbf{25.8}$  \\
Male (EN)                     & $19.7$  & $17.0$  & $17.9$           & $19.1$           & $19.1$           \\
Cello                         & $19.2$  & $17.8$  & $\mathbf{21.6}$  & $\mathbf{23.3}$  & $\mathbf{24.4}$  \\
Flute                         & $20.8$  & $16.2$  & $19.6$           & $\mathbf{23.3}$  & $\mathbf{23.8}$  \\
\midrule
\textbf{Mean (70 files)}  & $21.93$ & $18.86$ & $21.35$          & $\mathbf{22.44}$ & $\mathbf{23.18}$ \\
\bottomrule
\end{tabular}
\end{small}
\label{tab:blocksize_perfile}
\end{center}
\end{table}
The NMR results are similar (Table~\ref{tab:blocksize_perfile_nmr}): the $w=512$ MLP matches or beats the $w=2048$ baseline on $4$ of $7$ items, the linear predictor on $5$, and the CNN on $6$. \emph{Male (EN)} (speech) again resists all three, and \emph{soprano} is recovered only by the CNN; the MLP's remaining shortfall is \emph{glockenspiel}, consistent with its weakness on struck content noted earlier. Across all $70$ files, all three $w=512$ predictors beat the $w=2048$ baseline in mean NMR.

\begin{table}[t]
\caption{Mean NMR (dB) per file, $w=2048$ baseline against the four $w=512$ configurations. Bolded cells meet or beat the $w=2048$ baseline.}
\begin{center}
\begin{footnotesize}
\setlength{\tabcolsep}{3pt}
\begin{tabular}{lccccc}
\toprule
         & $w=2048$ & \multicolumn{4}{c}{$w=512$} \\
\cmidrule(lr){3-6}
File                          & Base     & Base    & Linear            & MLP               & CNN               \\
\midrule
Glockenspiel                  & $-21.3$  & $-19.1$ & $\mathbf{-21.8}$  & $-21.0$           & $\mathbf{-22.6}$  \\
Harpsichord                   & $-13.3$  & $-12.1$ & $\mathbf{-15.6}$  & $\mathbf{-14.9}$  & $\mathbf{-17.3}$  \\
Organ                         & $-10.8$  & $-9.3$  & $\mathbf{-13.2}$  & $\mathbf{-13.3}$  & $\mathbf{-15.0}$  \\
Soprano                       & $-17.0$  & $-12.9$ & $-15.7$           & $-16.4$           & $\mathbf{-17.8}$  \\
Male (EN)                     & $-15.1$  & $-12.1$ & $-13.6$           & $-14.0$           & $-14.4$           \\
Cello                         & $-14.1$  & $-11.3$ & $\mathbf{-14.5}$  & $\mathbf{-15.1}$  & $\mathbf{-16.6}$  \\
Flute                         & $-13.0$  & $-10.1$ & $\mathbf{-13.9}$  & $\mathbf{-14.4}$  & $\mathbf{-16.0}$  \\
\midrule
\textbf{Mean (70 files)}      & $-18.11$ & $-15.56$ & $\mathbf{-18.19}$  & $\mathbf{-18.23}$  & $\mathbf{-19.27}$ \\
\bottomrule
\end{tabular}
\end{footnotesize}
\label{tab:blocksize_perfile_nmr}
\end{center}
\end{table}

In summary, at $w=512$ our neural predictors match, and with the CNN exceed the coded SNR and NMR obtained by the non-predictive baseline after quadrupling the analysis-window length to $w=2048$, at substantially lower latency. This result favors pairing the shorter analysis window with a lightweight predictor over using the longer window.

\subsection{Listening Tests}
\label{sec:mushra}

\begin{figure*}[t]
\centering
\includegraphics[width=0.95\textwidth]{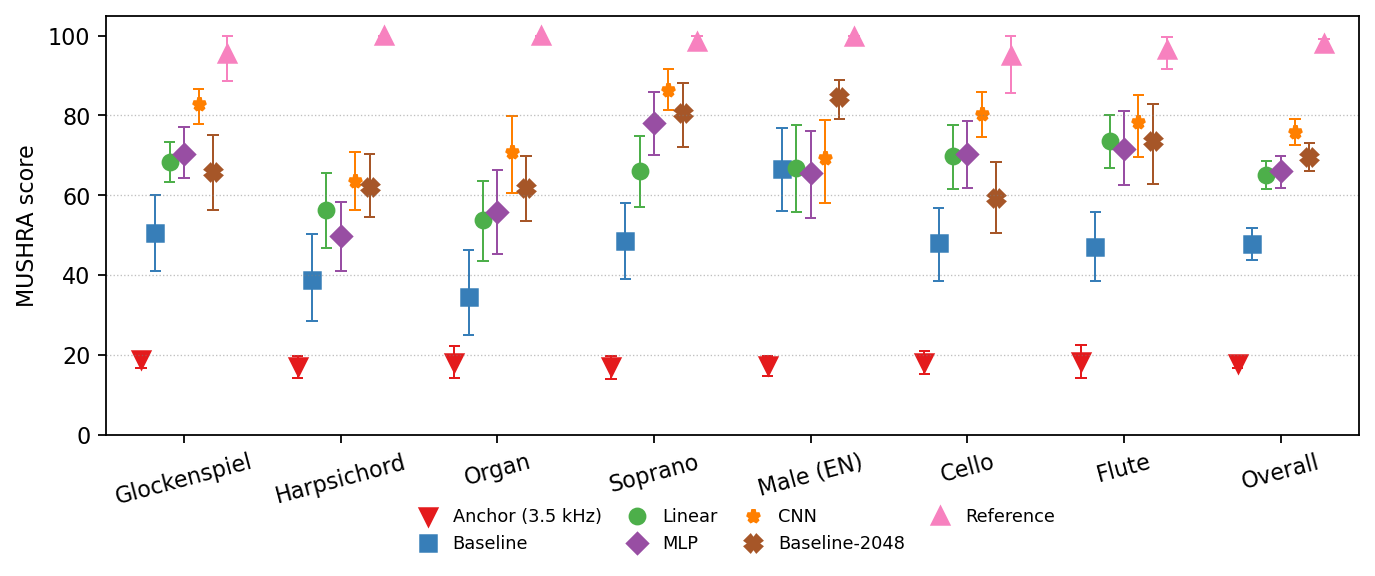}
\caption{Mean MUSHRA score by item (and overall) per condition, with $95\%$ confidence intervals and $N=17$ trained listeners.}
\label{fig:mushra_per_item}
\end{figure*}

Objective metrics alone do not fully capture perceptual quality, and listeners may be sensitive to artifacts that SNR and NMR miss. We conducted a subjective evaluation following ITU-R BS.1534 \citep{itur_bs1534_2015} (MUSHRA) recommendation. Each trial included the hidden reference, a $3.5$~kHz low-pass anchor, and five coded examples: the four $w=512$ codecs plus the non-predictive $w=2048$ baseline from Section~\ref{sec:blocksize}. All conditions were encoded at $\sim 110$~kbps ($\approx 2.5$~bits/sample; mono, duplicated to stereo for playback via high-quality headphones, $44.1$~kHz).\footnote{Audio samples and supplemental materials are available at:\linebreak \urlstyle{sf}\url{github.com/NiccoloAbate/neural-predictive-audio-coding}.} The $N=17$ trained listeners all passed standard post-screening \citep{itur_bs1534_2015}.

Figure~\ref{fig:mushra_per_item} shows per-item and overall mean MUSHRA scores with $95\%$ confidence intervals. Consistent with the objective results, the $w=512$ CNN is the clearest perceptual winner: it significantly outperforms the $w=512$ baseline ($\Delta = +28$ points), the linear predictor ($\Delta = +11$ points), and, corroborating Section~\ref{sec:blocksize}, the non-predictive $w=2048$ baseline ($\Delta = +6$ points). All three of these effects are significant at $p < 0.01$ (Holm-corrected paired $t$-tests). All three predictors significantly beat the $w=512$ baseline, though the linear and MLP conditions are statistically indistinguishable from each other.

\section{Profiling and Runtime Optimization}
\label{sec:profiling}

The quality and latency arguments of Sections~\ref{sec:objective} and~\ref{sec:blocksize} are only useful if the predictor is reasonably performant. Here, we profile the four codecs at $w=512$, and identify the dominant bottlenecks at encode and decode. We further explore three optimization targets: the encoder-side cost of conditional prediction, the CPU/GPU choice for sequential decoding, and prefetching the next prediction to hide its latency.

All measurements come from our research Python implementation, run on an Apple M4 Pro laptop. The two neural predictors run under ONNX Runtime (ORT) and so are close to production cost for the model inference, but the surrounding codec stages run as unoptimized NumPy code. A production implementation in a lower-level language would be substantially faster, so we focus our discussion on \emph{ratios} between configurations rather than absolute timings.

\subsection{Wall-Clock Cost}
\label{sec:wallclock}

The encoder does substantially more work per coefficient block than the decoder, since the quantization-aware gate (Eq.~\eqref{eq:gate}) evaluates whole-block quantize--dequantize round trips for both the original and residual candidates before final quantization of the mixed coefficient block. We therefore report encode and decode wall-clock separately.

\begin{table}[t]
\caption{Mean per-item encode/decode wall-clock (s), difference ($\Delta$) from baseline (s), and decoder real-time factor (RTF). CPU, single channel, 70 EBU SQAM files, $f_s=44.1$~kHz, $w=512$.}
\begin{center}
\begin{small}
\setlength{\tabcolsep}{4pt}
\begin{tabular}{lccccc}
\toprule
Mode     & Enc.~(s) & $\Delta$Enc.~ & Dec.~(s) & $\Delta$Dec. & Dec.~RTF \\
\midrule
Baseline & $12.7$ & ---      & $2.8$ & ---     & $0.05$ \\
Linear   & $22.0$ & $+9.3$  & $2.9$ & $+0.2$ & $0.06$ \\
MLP      & $23.1$ & $+10.4$ & $3.9$ & $+1.1$ & $0.08$ \\
CNN      & $24.4$ & $+11.6$ & $4.8$ & $+2.1$ & $0.10$ \\
\bottomrule
\end{tabular}
\end{small}
\label{tab:wallclock}
\end{center}
\end{table}

All four configurations decode faster than real time on average (Table~\ref{tab:wallclock}). Even the CNN has a decoder RTF of only $\approx 0.10$. The encoder shows the asymmetry clearly: enabling prediction nearly doubles encode wall-clock ($12.7 \to 24.4$~s for the CNN) but adds only $+2.1$~s at decode, with most of the extra encode cost going to the per-band gating decision rather than the predictor itself (Section~\ref{sec:gate_proxy}). %
The CNN decodes more slowly than the MLP ($4.8$ vs.\ $3.9$~s) despite being far smaller --- $0.13$M parameters ($0.5$~MB) against the MLP's $2.63$M ($10.5$~MB).

\subsection{CPU vs.\ GPU Inference} %
\label{sec:cpu_vs_gpu}

Neural inference is often performed on the GPU, but the codec is fundamentally sequential: each predicted block depends on the previously reconstructed one, so a single-channel decoder cannot batch blocks at all. The only parallelism within a single file is across its audio channels, so the batch size cannot exceed the channel count (two for stereo); larger batches require decoding several files at once. At batch~$1$, the dominant cost of a GPU forward pass is not compute but the host$\leftrightarrow$device round trip.

\begin{table}[t]
\caption{Per-call CNN predictor latency, isolated from the codec pipeline. ONNX Runtime on CPU vs.\ PyTorch MPS backend, mean over $300$ calls after warmup, at ($w=512$, $C=5$) and ($w=2048$, $C=3$).}
\begin{center}
\begin{small}
\setlength{\tabcolsep}{4pt}
\begin{tabular}{llcc}
\toprule
$w$ & Workload                               & CPU (ms) & MPS (ms) \\
\midrule
$512$  & batch~$1$ (single channel)          & $\mathbf{0.22}$ & $1.21$ \\
$512$  & batch~$2$ (stereo)                  & $\mathbf{0.74}$ & $1.17$ \\
$512$  & batch~$8$ (cross-file)              & $\mathbf{1.51}$ & $1.69$ \\
$512$  & batch~$32$ (cross-file)             & $3.99$ & $\mathbf{2.42}$ \\
$2048$ & batch~$1$ (single channel)          & $\mathbf{0.57}$ & $1.34$ \\
$2048$ & batch~$2$ (stereo)                  & $2.86$ & $\mathbf{1.70}$ \\
$2048$ & batch~$8$ (cross-file)              & $5.65$ & $\mathbf{2.07}$ \\
\bottomrule
\end{tabular}
\end{small}
\label{tab:cpuvsgpu}
\end{center}
\end{table}

At single-channel decode (batch~$1$) the CPU is significantly faster than the MPS backend at both block sizes (Table~\ref{tab:cpuvsgpu}): the host$\leftrightarrow$device round trip exceeds the entire CPU forward pass, so the GPU never amortizes its transfer cost. Batching shifts the balance, but the point at which the GPU overtakes the CPU varies: near batch~$32$ at $w=512$, but as early as stereo (batch~$2$) at $w=2048$, where the larger per-call work fills the GPU sooner. For single-channel sequential decode, then, the CPU is the right device at either block size, and the GPU pays off only once batching is available, which happens at smaller batches when the blocks are larger. Keeping the codec on the CPU also avoids competing with other GPU workloads and supports devices with no usable GPU stack, which is important for broad deployment.

\subsection{Conditional Prediction Gate Approximation}
\label{sec:gate_proxy}

As shown in Section~\ref{sec:wallclock}, the quantization-aware gate's whole-block candidate round trips dominate the predictive encoder's wall-clock. We therefore evaluate two cheaper approximations that estimate the gate's decision without these round trips. The \emph{band-energy} approximation selects prediction when the residual energy in a scale-factor band is lower than the corresponding energy in the original coefficients. The \emph{scaled-peak} approximation first multiplies the original and residual candidates by their respective power-of-two block-wide quantizer scales. It then selects prediction when the peak magnitude of the scaled residual in a band is lower than the corresponding peak of the scaled original coefficients. Unlike an unscaled peak comparison, this rule accounts for the different overall scale assigned to each candidate and therefore compares their band peaks at the input to the per-band quantizer. For reference, we also report an \emph{always-predict} trial that skips the gate entirely and uses prediction in every band regardless of per-band benefit.

\begin{table}[t]
\caption{Codec-aligned quality and gate-decision wall-clock under two approximations (energy, scaled peak) relative to the quant-aware gate, with an always-predict anchor for reference. CNN predictor at $w=512$, $C=5$, $2.0$~bits/sample; EBU SQAM set. ``Agree.'' is per-band mask agreement with the reference gate.}
\begin{center}
\begin{small}
\setlength{\tabcolsep}{3pt}
\begin{tabular}{lccccc}
\toprule
Gate          & $\Delta$SNR & $\Delta$NMR & Agree.\ & Pred. & Gate \\
              & (dB)        & (dB)        & (\%)    & (\%)  & (ms/block) \\
\midrule
Q-aware       & ---         & ---         & ---     & $36.9$ & $0.98$ \\
Energy        & $-0.52$     & $+0.48$     & $85.2$  & $36.6$ & $0.07$ \\
Scaled-peak   & $-3.63$     & $+1.71$     & $75.9$  & $21.8$ & $0.07$ \\
Always-pred.  & $-9.17$     & $+1.32$     & $36.9$  & $100$  & $0$    \\
\bottomrule
\end{tabular}
\end{small}
\label{tab:gate_proxy}
\end{center}
\end{table}

The band-energy approximation is competitive (Table~\ref{tab:gate_proxy}): for a small SNR/NMR cost it buys a $14\times$ reduction in gate-decision wall-clock, and it agrees with the reference gate on $85.2\%$ of per-band decisions. Only three files regress by more than $1$~dB SNR, the worst at $-1.59$~dB (kettle drums). The scaled-peak approximation is under-conservative without margin tuning and loses $3.6$~dB SNR. The always-predict trial confirms the gate is necessary: without it, in rare cases, prediction error may feed back through the reconstructed context and blow up (e.g. $-74.4$~dB SNR on celesta). We keep the quantization-aware gate in the headline results of Sections~\ref{sec:objective}--\ref{sec:blocksize}, but the energy approximation reduces encoder-side gating cost substantially for a modest quality loss, which makes it a useful encoder optimization where latency matters. The choice of gate lives entirely in the encoder: only the per-band mask is transmitted, so swapping gates is decoder-transparent and needs no bitstream or decoder changes.

We also implemented prefetching for the predictor. Because the prediction for block $n{+}1$ needs no input beyond the reconstruction of block $n$, the decoder can launch it as soon as that reconstruction finishes and overlap it with the current block's remaining CPU processing and, when streaming, with the wait for the next packet. In a sequential decoder this is the main way to hide the predictor's per-call latency at batch~$1$. Even unoptimized, our $w=512$ CNN has a decoder RTF of roughly $0.10$ on a CPU. These metrics are within range for practical deployment.

\section{Conclusion}

In this work, we investigated the novel use of small neural networks as black-box predictors in MDCT-based codec pipelines. We integrated a deterministic predictor into the residual-coding path of a standard MDCT coder while leaving the transform, psychoacoustic model, bit allocation, quantization, and synthesis stages unchanged. Because the predictor updates only from reconstructed coefficient blocks, the encoder and decoder stay synchronized without transmitting any predictor state, so the network behaves as a self-contained module that the rest of the codec need not understand.

Our results show that even very small neural networks can outperform existing predictive techniques, such as linear prediction. A CNN of only $0.13$M parameters ($0.5$~MB) slots into the existing predictive path and consistently outperforms the MLP, the NLMS linear predictor, and the non-predictive baseline in mean SNR, NMR, and subjective MUSHRA scores --- and does so with roughly $20\times$ fewer weights than the MLP, showing structural alignment winning out over raw capacity.

At $w=512$ the neural predictors recover, and the CNN exceeds, the coded quality of the non-predictive $w=2048$ configuration at substantially lower algorithmic delay. Where low-delay coders trade frequency resolution for delay by shortening the analysis window, our results point to a complementary adaptation: keep the short window and restore the lost quality with a small predictor. The runtime cost of adding the network is reasonable. Decoder RTF is about $0.10$ on a CPU, with no GPU required.

There are several natural directions for future research. Domain-specific predictors (e.g. for speech, tonal music, and percussion) could be signaled and switched at the block level. Alternatively, while we kept the models small by design to keep decode cheap, scaling them up in capacity and training data may prove fruitful. Larger predictors would raise the inference compute per block while likely improving prediction gain and coded quality, revealing a tradeoff between decode-time cost and codec efficiency that this work explores only at the smaller-model end. Architectures that pair the CNN's structural priors with longer-range modeling, such as a Conformer, are also a reasonable next step. Prediction-aware bit allocation could exploit the residual statistics to recover more of the raw prediction gain in coded SNR. Bounding the error propagation introduced by sequentially dependent predictions, for instance with periodic non-predictive refresh blocks, would also be important for streaming applications.

\end{document}